\documentclass[conference]{IEEEtran}
\IEEEoverridecommandlockouts
\usepackage{cite}
\usepackage{amsmath,amssymb,amsfonts}
\usepackage{algorithmic}
\usepackage{graphicx}
\usepackage{textcomp}
\usepackage{xcolor}
\usepackage{subfig}
\usepackage{booktabs}
\usepackage{import}
\usepackage{color}
\usepackage[citebordercolor=blue]{hyperref}
\usepackage[capitalise]{cleveref}
\usepackage{multirow}
\usepackage{array}
\usepackage[font=footnotesize]{caption}
\usepackage[inline]{enumitem}
\usepackage{makecell}
\usepackage{siunitx}
\usepackage{caption}
\usepackage{float}

\usepackage{tikz}
\usetikzlibrary{arrows.meta}
\usepackage[capitalise]{cleveref}
\crefname{figure}{Fig.}{Figs.}
\Crefname{figure}{Fig.}{Figs.}

\AtBeginDocument{%
  \setlength{\abovedisplayskip}{3pt plus 1pt minus 1pt}%
  \setlength{\belowdisplayskip}{3pt plus 1pt minus 1pt}%
  \setlength{\abovedisplayshortskip}{2pt plus 1pt minus 1pt}%
  \setlength{\belowdisplayshortskip}{2pt plus 1pt minus 1pt}%
}

\begin{document}
\bstctlcite{BSTcontrol}

\title{RAMA: Retrieval-Augmented Multi-Agent Routing with Episodic Memory for LEO Satellite Networks}

\author{\IEEEauthorblockN{Xin~Ming~Li\IEEEauthorrefmark{1}\IEEEauthorrefmark{4}~and~Peng~Hu\IEEEauthorrefmark{1}\IEEEauthorrefmark{3}\IEEEauthorrefmark{4}
\IEEEauthorblockA{
    \IEEEauthorrefmark{1}Centre for Computational Mathematics, University of Waterloo, Waterloo, Canada}
\IEEEauthorblockA{
    \IEEEauthorrefmark{3}Cheriton School of Computer Science, University of Waterloo, Waterloo, Canada}
\IEEEauthorblockA{
    \IEEEauthorrefmark{4}Dept. of Electrical and Computer Engineering, University of Manitoba, Winnipeg, Canada}
    xm5li@uwaterloo.ca, {peng.hu@umanitoba.ca}
}
\thanks{We acknowledge the support of the Natural Sciences and Engineering Research Council of Canada (NSERC), [funding reference number RGPIN-2022-03364]. The authors thank Ken Cao and Mahyar Tajeri for their contributions to research discussions and assistance with experiments.
}
}

\maketitle

\begin{abstract}
Low-Earth orbit (LEO) satellite networks present a challenging routing environment due to rapidly changing topology, traffic, and link conditions. 
Existing deep reinforcement learning (DRL) routing methods typically rely on current observations and information encoded in network weights, limiting their ability to exploit successful past experiences.
We propose RAMA, a retrieval-augmented multi-agent DRL framework with Approximate Information State (AIS), in which each satellite combines its local observation, recent routing history, neighboring state information, and retrieved entries from a per-satellite episodic memory. 
Retrieved experiences are integrated through a Transformer-based decision network trained with AIS-based auxiliary objectives. 
With weights frozen in a 140-satellite Kepler constellation, RAMA reduces mean end-to-end latency by 13.4\% and $99^\text{th}$-percentile latency by 66.7\% relative to the state-of-the-art double deep Q-network (DDQN) routing method, mainly by avoiding long detours.
Updated online, RAMA remains stable while DDQN degrades. 
On the 1,584-satellite Starlink shell 1, the reduction in mean latency grows to 93.2\%, and updating RAMA online lowers its mean latency by a further 25.5\%. These results demonstrate the effectiveness of retrieval-augmented memory for scalable decentralized routing in LEO satellite networks.
\end{abstract}


\section{Introduction}

As LEO constellations scale from hundreds to thousands of satellites, maintaining an accurate global network state becomes increasingly difficult. Routing decisions must often be made using only local information, while queue occupancies, traffic demands, and link availability evolve continuously. Under these conditions, shortest-path routes computed from stale information may quickly become suboptimal, resulting in congestion, increased latency, and poor resilience to failures \cite{hu2024}.

Recent deep reinforcement learning (DRL)-based routing approaches \cite{lozanocuadra2024continualdeepreinforcementlearning,chen2025transformermix,lyu2024constrained} improve adaptability by learning forwarding policies directly from interaction with the network. However, most existing methods make decisions solely from the current observation or a compact recurrent state. Consequently, valuable information from previous routing experiences may be forgotten, even when similar traffic and topological situations recur.

LEO constellations exhibit strong periodicity because satellites repeatedly revisit similar orbital positions. Likewise, many traffic patterns recur at daily or shorter timescales. Therefore, forwarding decisions that previously led to low-delay packet delivery can provide useful guidance when similar conditions are encountered again. State-of-the-art multi-agent DRL (MADRL) schemes \cite{lozanocuadra2024continualdeepreinforcementlearning, chen2025transformermix} do not explicitly retain or retrieve such experiences.

To address this limitation, we propose Retrieval-Augmented Memory with Approximate Information State (AIS), called RAMA, a retrieval-enhanced MADRL framework for LEO packet routing. Each satellite maintains a lightweight episodic memory containing previous forwarding experiences together with their realized outcomes. When making a routing decision, the satellite retrieves experiences most similar to the current state and combines them with local observations, neighboring satellite information, and recent routing history through a Transformer-based decision network. Unlike existing LEO routing approaches that encode experience only through network weights or recurrent hidden states, RAMA explicitly stores and retrieves routing experiences at inference time. This enables adaptation through memory even when model parameters remain fixed after deployment. The main contributions are summarized as follows:

\begin{itemize}
    \item We propose RAMA that combines local observations, routing history, neighbor states, and a per-satellite episodic memory.
    \item We integrate retrieval-based memory with the AIS learning framework, enabling learned representation to capture both reward- and dynamics-relevant information.
    \item We extensively evaluate RAMA against double deep Q-network (DDQN) and Dijkstra-based shortest-path routing under both frozen and online adaptation settings.
    \item We provide packet-level insights into RAMA's performance gains and the trade-offs relative to shortest-path routing.
\end{itemize}

The rest of the paper is organized as follows.
\Cref{sec:related} discusses the related work.
\Cref{sec:method} presents the system model, the RAMA architecture and its training process.
\Cref{sec:results} presents the experimental results and a discussion.
\Cref{sec:conclusion} concludes the paper and outlines future work.

\section{Related Work}
\label{sec:related}

\subsection{Routing in LEO Constellations}
Existing LEO routing methods can be broadly categorized into shortest-path approaches that rely on explicit network models and DRL-based approaches that learn forwarding policies from interaction with the environment. While DRL methods improve adaptability, historical information is typically encoded in recurrent states, topology histories, or continually updated network parameters. As a result, successful forwarding experiences cannot be explicitly retrieved when similar network conditions recur. This limitation is particularly relevant in LEO constellations, where orbital geometry and traffic patterns exhibit recurring structure. 

Although Dijkstra-based shortest-path algorithm remains a widely used routing method in LEO networks~\cite{celikbilek2022survey}, recent works have explored DRL-based routing policies. Lozano-Cuadra \textit{et al.}~\cite{lozano2024opensource} released an event-based simulator of LEO packet routing that supports shortest-path, tabular Q-routing and MADRL policies. Their MADRL routing method~\cite{lozanocuadra2024continualdeepreinforcementlearning} models each satellite as a DDQN agent that selects the next hop using its position, the destination and neighboring queue states. 
The policy makes decisions from the current observation alone and adapts online solely through updates to its network weights. Chen \textit{et al.}~\cite{chen2025transformermix} proposed a Transformer-MIX learning approach in which each agent combines a gated recurrent unit with self-attention over its current observation, previous action and agent identifier. In TransfQMix~\cite{TransQMix23}, a Transformer generates the weights of the mixing network used during centralized training. Lyu \textit{et al.}~\cite{lyu2024constrained} proposed a constrained multi-agent reinforcement learning dynamic routing (CMADR) for integrated satellite-terrestrial networks, where satellites and ground stations (GSs) are jointly optimized to minimize packet delay subject to energy and packet-loss constraints using a Lagrangian max-min formulation with centralized reward and cost critics. He \textit{et al.}~\cite{he2026gnnasssp} instead computed end-to-end routes from a network-wide view by predicting link availability from recent topology snapshots with a spatiotemporal Transformer and learning congestion-aware edge weights for shortest-path search using a graph attention network.
The first two choose joint actions in synchronized time steps, which per-packet, event-driven methods do not provide, and the third relies on a network-wide view. Although these methods exploit historical information, they do so through recurrent hidden states~\cite{chen2025transformermix}, topology histories~\cite{he2026gnnasssp}, or online parameter updates~\cite{lozanocuadra2024continualdeepreinforcementlearning}. None maintains a per-satellite memory of forwarding decisions together with their realized end-to-end delivery outcomes.

\subsection{Memory in Reinforcement Learning}
Memory mechanisms in reinforcement learning generally fall into two categories: parametric memory, in which past information is compressed into network parameters or recurrent hidden states, and non-parametric memory, in which individual experiences are explicitly stored and later retrieved. Both have been used to improve decision making under partial observability and non-stationarity. A representative example of parametric memory is the deep recurrent Q-network (DRQN) of Hausknecht and Stone~\cite{hausknecht2015drqn}, which replaces a fully connected layer in a deep Q-network with a long short-term memory module. The recurrent hidden state summarizes recent observations, enabling decisions based on short histories rather than a single observation. 

Non-parametric approaches instead retain individual experiences. Blundell \textit{et al.}~\cite{blundell2016mfec} proposed model-free episodic control, which stores the best return observed for each state-action pair and estimates values from nearby stored experiences. Pritzel \textit{et al.}~\cite{pritzel2017nec} extended this idea through neural episodic control, which learns state embeddings and stores them in differentiable memory structures indexed by action. In both methods, retrieved memories directly produce value estimates, and both were developed for single-agent gaming environments. More recently, retrieval has been combined with large external experience repositories. Humphreys \textit{et al.}~\cite{humphreys2022r2a} conditioned a Go-playing agent on similar states retrieved from a large corpus of expert games. Schmied \textit{et al.}~\cite{schmied2024radt} proposed the Retrieval-Augmented Decision Transformer, which retrieves sub-trajectories from previously collected experience and incorporates them into in-context decision making. These methods demonstrate the value of explicit experience retrieval but focus on single-agent settings and are not designed for decentralized packet routing in large-scale satellite networks.

\subsection{Novelty of RAMA}
\label{subsec:novelty}
RAMA differs from prior routing and memory-based approaches in four key respects. First, each satellite maintains a compact private memory of forwarding experiences, retrospectively labeled with realized delivery outcomes and accessed through similarity-based retrieval. Second, retrieved memories are integrated with current observations, recent routing history, and neighbor information through a Transformer-based decision network rather than being used directly for value estimation. Third, the policy is trained with auxiliary objectives from the AIS framework, which, to the best of our knowledge, have not previously been applied to packet routing or combined with retrieval-augmented memory. Finally, the memory remains active after training, enabling deployed agents to continuously record and retrieve new experiences even when policy parameters are frozen. To our knowledge, RAMA is the first LEO routing framework to unify decentralized packet forwarding, per-satellite retrieval-augmented episodic memory, and AIS-based representation learning within a single multi-agent reinforcement learning architecture.

\section{Proposed Method}
\label{sec:method}

\subsection{System Model}
\label{subsec:system}

The network formulation in~\cite{lozanocuadra2024continualdeepreinforcementlearning} is adopted.
GSs connect users on the ground to the constellation, each through a link to its nearest satellite, and every packet travels from a source GS to a destination GS across inter-satellite links.
A packet of $B$ bits crossing the link from node $i$ to node $j$, where a node is a satellite or a GS, incurs the latency
\begin{equation}
T(i,j)=t_q(i)+\frac{B}{R(i,j)}+\frac{\|ij\|}{c},
\label{eq:hop-latency}
\end{equation}
where $t_q(i)$ is its wait time in node $i$'s queue, $R(i,j)$ is the link rate, $\|ij\|$ is the slant range between nodes $i$ and $j$, and $c$ is the speed of light.
The end-to-end latency $T_p$ of packet $p$ is the sum of \eqref{eq:hop-latency} over every link it traverses.

For a packet $p$ at the satellite node $i$ making its $n^{\text{th}}$ hop, with $d$ the satellite linked to $p$'s destination GS, the agent on $i$ selects the next hop $a_{i,n}\in\mathcal{A}_i=\{\mathrm{U}_i,\mathrm{D}_i,\mathrm{R}_i,\mathrm{L}_i\}$ of $p$.
The neighbors of satellite $i$ above and below it in its orbital plane are $\mathrm{U}_i$ and $\mathrm{D}_i$, respectively.
Similarly, the neighbors of $i$ to its right and left in the adjacent planes are $\mathrm{R}_i$ and $\mathrm{L}_i$.
The permitted set of actions $\mathcal{A}_{i,n}$ is the subset of the action space $\mathcal{A}_i$ containing the neighbors of $i$ except, where possible, the one $p$ came from, and excluding any neighbor that $p$ has visited five times or more. 
This is a loop guard to prevent cycling.
If $a_{i,n}\notin\mathcal{A}_{i,n}$, then $a_{i,n}$ is removed from ${\mathcal{A}_{i}}$ and the agent makes another decision.
If $d$ is a neighbor of $i$, then the agent is bypassed and $p$ is routed to $d$.

Inspired by~\cite{lozanocuadra2024continualdeepreinforcementlearning}, satellite $i$ observes a 31-dimensional vector
\begin{align}
o_n&=\big[v_d,v_{\mathrm{R}_i},v_{\mathrm{L}_i},\mathbf{b}_{\mathrm{U}_i},\mathbf{b}_{\mathrm{D}_i},\mathbf{b}_{\mathrm{R}_i},\mathbf{b}_{\mathrm{L}_i},\bar{\phi}_i,\bar{\lambda}_i,\bar{\phi}_d,\bar{\lambda}_d\big],\nonumber\\
\mathbf{b}_j&=\big[C_{j,\mathrm{U}_j},C_{j,\mathrm{D}_j},C_{j,\mathrm{R}_j},C_{j,\mathrm{L}_j},\bar{\phi}_j,\bar{\lambda}_j\big],\quad j\in\mathcal{A}_i,
\label{eq:obs}
\end{align}
where $\mathbf{b}_j$ holds the position of neighbor $j$ and the congestion scores $C_{j,k}$ of $j$'s queues toward $j$'s neighbor $k$.
Positions are latitudes $\phi$ and longitudes $\lambda$ in whole degrees, normalized as $\bar{\phi}=(\phi+90)/\sigma$ and $\bar{\lambda}=(\lambda+180)/\sigma$ with $\sigma=20$.
The direction score of satellite $j$ is $v_j=-\sin\psi_j$, where $\psi_j$ is $j$'s angle within its orbital plane measured from the plane's northernmost point, so $v_j$ is positive while $j$ moves north and negative while it moves south.
The congestion score of satellite $j$'s queue for $j$'s neighbor $k$ is
\begin{equation}
C_{j,k}=\min\Big(C^\ast,\Big\lfloor C^\ast\,\frac{\log_{10}(q_{t}+1)}{\log_{10}q_{\max}}\Big\rfloor\Big),
\label{eq:queue-score}
\end{equation}
where $C^\ast=10$ is the maximum congestion value, $q_{\max}$ is the maximum queue size, and $q_t$ is the queue length at time $t$~\cite{lozanocuadra2024continualdeepreinforcementlearning}.
A missing neighbor has every congestion score set to infinite and coordinates at 0.

The reward for an agent's decision $a_{i,n}$ to forward packet $p$ from $i$ to $j$ is
\begin{align}
r_n={}&w_1\big(1-10^{\,t_q(i)}\big)+w_2\,\frac{\|id\|-\|jd\|-\|ij\|/w_4}{d_{\max}}\nonumber\\
&+w_3\big(r_l\,\mathbb{I}[j\in\mathcal{P}]+r_d\,\mathbb{I}[j=d]+r_u\,\mathbb{I}[a_{i,n}\notin\mathcal{A}_{i,n}]\big),
\label{eq:reward}
\end{align}
where $\mathbb{I}[\cdot]$ is the indicator, $d_{\max}$ is the longest inter-satellite link, and $\mathcal{P}$ is the set of satellites the packet visited before $i$. 
The weights are $w_1=w_2=20$, $w_3=1$ and $w_4=5$.
The term $r_l=-10$ penalizes a revisit, $r_d=50$ rewards reaching $d$, and $r_u=-10$ penalizes an attempted hop outside $\mathcal{A}_{i,n}$~\cite{lozanocuadra2024continualdeepreinforcementlearning}.

\subsection{RAMA Architecture}
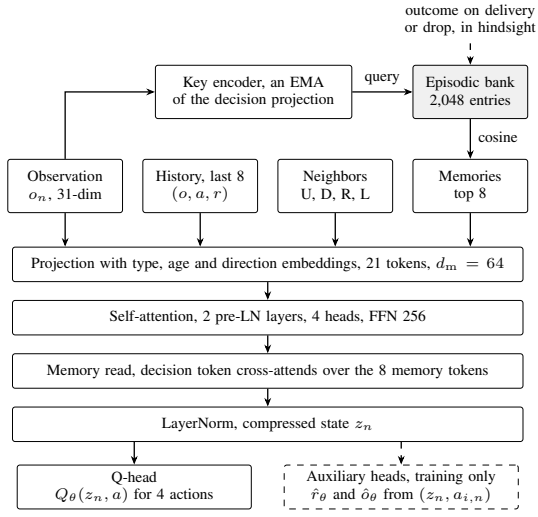
\begin{figure}[!t]
\centering
\scalebox{0.83}{
\begin{tikzpicture}[
  x=1mm, y=1mm,
  font=\scriptsize,
  box/.style={draw, rounded corners=1.2pt, align=center, inner sep=1.8pt},
  inp/.style={box, text width=17mm, minimum height=8.5mm},
  stage/.style={box, text width=80mm, minimum height=5.5mm},
  arr/.style={-{Stealth[length=1.4mm,width=1.1mm]}, semithick},
  darr/.style={arr, dashed}
]
\node[box, text width=30mm, minimum height=8.5mm] (key) at (40,15)
  {Key encoder, an EMA\\of the decision projection};
\node[box, text width=17mm, minimum height=8.5mm, fill=black!6] (bank) at (74.5,15)
  {Episodic bank\\2{,}048 entries};
\node[align=center] (out) at (74.5,26.5) {outcome on delivery\\or drop, in hindsight};
\node[inp] (obs)  at (10,0)   {Observation\\$o_n$, 31-dim};
\node[inp] (hist) at (31.5,0) {History, last 8\\$(o,a,r)$};
\node[inp] (nei)  at (53,0)   {Neighbors\\U, D, R, L};
\node[inp] (mem)  at (74.5,0) {Memories\\top 8};
\node[stage] (emb) at (42.25,-12.5) {Projection with type, age and direction embeddings, 21 tokens, $d_{\mathrm{m}}=64$};
\node[stage] (sa)  at (42.25,-21)   {Self-attention, 2 pre-LN layers, 4 heads, FFN 256};
\node[stage] (ca)  at (42.25,-29.5) {Memory read, decision token cross-attends over the 8 memory tokens};
\node[stage] (z)   at (42.25,-38)   {LayerNorm, compressed state $z_n$};
\node[box, text width=36mm, minimum height=7mm] (q) at (21,-48)
  {Q-head\\$Q_\theta(z_n,a)$ for 4 actions};
\node[box, text width=36mm, minimum height=7mm, dashed] (aux) at (63.5,-48)
    {Auxiliary heads, training only\\$\hat r_\theta$ and $\hat o_\theta$ from $(z_n,a_{i,n})$};
\draw[arr] (obs.north) |- (key.west);
\draw[arr] (key.east) -- node[above]{query} (bank.west);
\draw[arr] (bank.south) -- node[right, pos=0.45]{cosine} (mem.north);
\draw[darr] (out.south) -- (bank.north);
\foreach \n in {obs,hist,nei,mem} \draw[arr] (\n.south) -- (\n.south |- emb.north);
\draw[arr] (emb) -- (sa);
\draw[arr] (sa) -- (ca);
\draw[arr] (ca) -- (z);
\draw[arr] (z.south -| q.north) -- (q.north);
\draw[darr] (z.south -| aux.north) -- (aux.north);
\end{tikzpicture}
}
\caption{RAMA's decision network at one satellite.
Arrows show forward pass at decision time, except the write into the bank, which happens in hindsight when the packet is delivered/dropped.
Auxiliary heads are used only in training.}
\label{fig:arch}\vspace{-20pt}
\end{figure}

\Cref{fig:arch} shows RAMA's decision network.
At each decision the agent on satellite $i$ forms 21 tokens of width $d_{\mathrm{m}}=64$, one decision token, eight history tokens, four neighbor tokens and eight memory tokens,
\begin{align}
x^{\mathrm{dec}}&=W^{\mathrm{dec}}o_n+e_1,\nonumber\\
x^{\mathrm{h}}_\ell&=W^{\mathrm{h}}\big[o^{(\ell)};\mathbf{a}^{(\ell)};\tanh(r^{(\ell)}/50)\big]+e_2+e^{\mathrm{age}}_\ell,\nonumber\\
x^{\mathrm{nei}}_j&=W^{\mathrm{nei}}\big[\mathbf{b}_j;\beta_j\big]+e_3+e^{\mathrm{dir}}_j,\nonumber\\
x^{\mathrm{mem}}_m&=W^{\mathrm{mem}}\big[\kappa_m;\mathbf{a}_m;\tanh(g_m/0.1);\varsigma_m;s_m\big]+e_4,
\label{eq:tokens}
\end{align}
where $W^{\mathrm{dec}}$, $W^{\mathrm{h}}$, $W^{\mathrm{nei}}$ and $W^{\mathrm{mem}}$ are learned affine maps of the concatenated inputs,  $\mathbf{a}$ is a one-hot action, and $e_1,\dots,e_4$, $e^{\mathrm{age}}_\ell$ and $e^{\mathrm{dir}}_j$ are learned type, age and direction embeddings.
History token $x^{\mathrm{h}}_\ell$, $\ell=1,\dots,8$, holds the observation $o^{(\ell)}$, action $\mathbf{a}^{(\ell)}$ and reward $r^{(\ell)}$ of $i$'s $\ell^{\text{th}}$ most recent decision.
Neighbor token $x^{\mathrm{nei}}_j$ carries the flag $\beta_j$, which is 1 if neighbor $j$ is linked.
Satellite $i$ retrieves from its episodic bank the set $\mathcal{M}_n$ of the eight entries whose keys are most similar to the key of $o_n$.
Memory token $x^{\mathrm{mem}}_m$, $m\in\mathcal{M}_n$, represents one of these entries, where $\kappa_m$ is the key of the observation at the past decision the entry records, $\mathbf{a}_m$ is the action chosen at that decision, $g_m$ is the realized delay to the destination, $\varsigma_m$ measures the time since the entry was written, and $s_m$ is the cosine similarity between $\kappa_m$ and the key of $o_n$.
These are defined formally in \eqref{eq:bank} to \eqref{eq:key}.

Two pre-layer-norm Transformer blocks update the token matrix $X\in\mathbb{R}^{21\times d_{\mathrm{m}}}$, whose rows are the tokens of \eqref{eq:tokens},
\begin{align}
X&\leftarrow X+\mathrm{MHA}\big(\mathrm{LN}(X)\big),\nonumber\\
X&\leftarrow X+\mathrm{FFN}\big(\mathrm{LN}(X)\big),\nonumber\\
\mathrm{head}&=\mathrm{softmax}\big(QK^{\top}/\sqrt{16}+M\big)\,V,
\label{eq:block}
\end{align}
where LN is layer normalization, MHA is multi-head attention with four heads, $Q$, $K$ and $V$ are a head's learned query, key and value projections of $\mathrm{LN}(X)$ to width 16, the mask $M$ is $-\infty$ on empty history and memory slots, and FFN is a feed-forward network of width 256 with GELU activation.
After the second Transformer block, $\tilde{x}^{\mathrm{dec}}$ and $\tilde{X}^{\mathrm{mem}}$ denote the decision token and the eight memory tokens.
The blocks in \eqref{eq:block} mix all 21 tokens.
The decision token then attends to the memory tokens only, which gives the retrieved entries a direct route into the compressed state
\begin{equation}
z_n=\mathrm{LN}\Big(\tilde{x}^{\mathrm{dec}}+\mathrm{MHA}\big(\mathrm{LN}(\tilde{x}^{\mathrm{dec}}),\mathrm{LN}(\tilde{X}^{\mathrm{mem}})\big)\Big),
\label{eq:z}
\end{equation}
where this attention step takes its queries from $\mathrm{LN}(\tilde{x}^{\mathrm{dec}})$ and its keys and values from $\mathrm{LN}(\tilde{X}^{\mathrm{mem}})$.
While $i$'s bank is empty there is nothing to attend to, so the MHA term is dropped and $z_n=\mathrm{LN}(\tilde{x}^{\mathrm{dec}})$.
A linear head on the compressed state $z_n\in\mathbb{R}^{d_{\mathrm{m}}}$ gives the action values, and RAMA acts greedily within the permitted set, so it never incurs $r_u$,
\begin{equation}
Q_\theta(z_n,a)=\mathbf{u}_a^{\top}z_n+u^0_a,\quad a_{i,n}=\arg\max_{a\in\mathcal{A}_{i,n}}Q_\theta(z_n,a),
\label{eq:q}
\end{equation}
where $\mathbf{u}_a$ and $u^0_a$ are the head's weights and bias for action $a$, and $\theta$ denotes the network's parameters.

Each satellite $i$ keeps its own bank of keys $\kappa_m$, actions $a_m$, outcomes $g_m$ and write times $\tau_m$,
\begin{equation}
\mathcal{B}^i=\{(\kappa_m,a_m,g_m,\tau_m)\}_{m=1}^{|\mathcal{B}^i|},\qquad|\mathcal{B}^i|\le 2{,}048,
\label{eq:bank}
\end{equation}
and the oldest entry is overwritten when $\mathcal{B}^i$ is full. 
Each entry is written into the deciding satellite's bank on delivery or drop of the packet, with the outcome
\begin{equation}
g=\begin{cases}t_{\mathrm{del}}-t_{\mathrm{dec}}, & \text{if the packet is delivered},\\ 0.6~\mathrm{s}, & \text{if it is dropped},\end{cases}
\label{eq:outcome}
\end{equation}
where $t_{\mathrm{dec}}$ is the decision time and $t_{\mathrm{del}}$ is the delivery time.
Keys and queries are normalized outputs of an affine map $W^{\mathrm{key}}$ that has the same shape as $W^{\mathrm{dec}}$ but is not trained by gradient descent.
Instead, after every gradient step it moves a fraction $\eta=0.01$ of the way toward the current $W^{\mathrm{dec}}$,
\begin{equation}
\kappa(o)=\frac{W^{\mathrm{key}}o}{\|W^{\mathrm{key}}o\|_2},\qquad W^{\mathrm{key}}\leftarrow(1-\eta)\,W^{\mathrm{key}}+\eta\,W^{\mathrm{dec}},
\label{eq:key}
\end{equation}
where the update averages the two maps' weights and biases, so $W^{\mathrm{key}}$ is a slowly moving average of $W^{\mathrm{dec}}$ and keys written to the bank earlier stay comparable with queries computed now.
Retrieval returns the set $\mathcal{M}_n$ of the eight entries with the largest $s_m=\kappa_m^{\top}\kappa(o_n)$, which is the cosine similarity between $\kappa_m$ and the current query $\kappa(o_n)$ because both have unit length.
Each retrieved entry enters \eqref{eq:tokens} with staleness ${\varsigma_m=\min\{t_{\mathrm{now}}-\tau_m,\,1\}}$, where $t_{\mathrm{now}}$ is the current time and all times are in seconds.

\subsection{Training}

RAMA is trained with decaying random exploration using a single network shared by all satellites, so the index $i$ is dropped below.
During training, the loop guard introduced in \Cref{subsec:system} is not applied.
The $n^{\text{th}}$ RAMA decision gives a transition $(z_n,a_n,r_n,z_{n+1},\chi_n)$, where $\chi_n=1$ if the hop reaches $d$.
The transition is stored in a replay buffer when the packet's next decision is made or the packet reaches $d$.
RAMA is trained using double Q-learning with unavailable neighbors masked out,
\begin{equation}
y_n=r_n+\gamma(1-\chi_n)\,Q_{\theta^-}\Big(z^-_{n+1},\arg\max_{a\in\mathcal{A}^{\mathrm{link}}_{n+1}}Q_\theta(z_{n+1},a)\Big),
\label{eq:target}
\end{equation}
where $z_{n+1}$ and $z^-_{n+1}$ are computed by the networks with parameters $\theta$ and $\theta^-$, respectively.
The set $\mathcal{A}^{\mathrm{link}}_{n+1}$ holds the linked neighbors at the next hop, and the target network's parameters $\theta^-$ are copied from $\theta$ every 1,000 routing decisions.
Two auxiliary heads predict the reward $\hat{r}_\theta$ and the next observation $\hat{o}_\theta$ from $z_n$ and the action, and each gradient step minimizes over a minibatch of 32 replayed transitions
\begin{align}
\mathcal{L}(\theta)=\mathbb{E}\Big[&h\big(Q_\theta(z_n,a_n)-y_n\big)+\lambda_r\big(\hat{r}_\theta-r_n\big)^2\nonumber\\
&+\frac{\lambda_o}{31}\big\|\hat{o}_\theta-o_{n+1}\big\|_2^2\Big],
\label{eq:loss}
\end{align}
where $h$ is the Huber loss and $\lambda_r=\lambda_o=0.5$.

The auxiliary heads follow the AIS framework of Subramanian and Mahajan~\cite{subramanian2019ais}, which Subramanian \textit{et al.}~\cite{jayakumar2022approximateinformationstate} extended with integral probability metrics and reinforcement learning losses, and Kao and Subramanian~\cite{Kao2022AISinMARL} extended to multi-agent settings.
A compression $\hat{\sigma}$ of the history $H_t$, with a reward predictor $\hat{r}$ and a kernel $\hat{P}$ that predicts the next compressed state, is an $(\varepsilon,\delta)$-AIS if for every $H_t$ and $a_t$
\begin{align}
\big|\mathbb{E}[r_t\mid H_t,a_t]-\hat{r}\big(\hat{\sigma}(H_t),a_t\big)\big|&\le\varepsilon,\nonumber\\
d_{\mathcal{F}}\big(\mu_t,\hat{P}(\cdot\mid\hat{\sigma}(H_t),a_t)\big)&\le\delta,
\label{eq:ais}
\end{align}
where $\mu_t$ is the distribution of the next compressed state $\hat{\sigma}(H_{t+1})$ given $(H_t,a_t)$ and $d_{\mathcal{F}}$ is the integral probability metric generated by a function class $\mathcal{F}$.
Acting greedily on the fixed point $\hat{Q}$ of the Bellman equation built from $\hat{r}$ and $\hat{P}$ then loses at most $2\alpha$ in value~\cite[Thm.~27]{jayakumar2022approximateinformationstate}, with
\begin{equation}
\alpha=\frac{\varepsilon+\gamma\,\rho_{\mathcal{F}}(\hat{V})\,\delta}{1-\gamma},
\label{eq:alpha}
\end{equation}
where $\hat{V}=\max_a\hat{Q}$ and $\rho_{\mathcal{F}}(\hat{V})$ is the Minkowski functional of $\hat{V}$ with respect to $\mathcal{F}$.
The reward and observation terms of \eqref{eq:loss} are the framework's surrogate losses for $\varepsilon$ and, through the next observation, for $\delta$~\cite{jayakumar2022approximateinformationstate}.

\section{Experimental Results}
\label{sec:results}

We evaluate RAMA on two representative LEO satellite constellations derived from FCC filings: Kepler, with 140 satellites in 7 planes at 600 km altitude and 98.6° inclination (Walker Star), and Starlink shell 1, with 1,584 satellites in 72 planes at 550 km altitude and 53° inclination (Walker Delta).
We extend the simulator~\cite{lozano2024opensource} with the RAMA agents and the direction scores.
Eight GSs exchange traffic through the constellation.
They are in Malaga, Los Angeles, Nemea, Nuuk, Bangalore, Mauritius, Vard{\o} and the Azores, which are locations from the existing Kongsberg Satellite Services (KSAT) GS network.
Each GS generates traffic using a Poisson process in proportion to the surrounding population and splits it equally among the other GSs.
Bangalore and Malaga having the largest populations generate the most traffic, and Nuuk the least.

RAMA and DDQN are evaluated in both frozen and online settings, initialized from the same trained model. In the frozen setting, network parameters remain fixed; however, RAMA continues to retrieve and update its episodic memory. In the online setting, each satellite independently updates its local model using replay-buffer training. Performance is measured by end-to-end latency, packets exceeding 0.5 s latency, mean hop count, and satellite revisit rate. A route is classified as long if its hop count exceeds twice Dijkstra's median hop count (14 on Kepler and 24 on Starlink).

We compare RAMA with Dijkstra and DDQN. Dijkstra computes minimum-hop routes using the exact global topology at each update and serves as an idealized benchmark with full network knowledge. DDQN ~\cite{lozanocuadra2024continualdeepreinforcementlearning}, a state-of-the-art decentralized LEO routing method, is evaluated under the same topology, traffic, action space, and reward settings as RAMA. Unlike RAMA, DDQN is a memoryless three-layer feed-forward network that bases decisions solely on the current observation, making it a natural decentralized baseline for assessing the benefits of retrieval-augmented memory.

\subsection{Performance Comparison}
\label{sec:performance}

\begin{table}[!t]
\renewcommand{\arraystretch}{1}
\setlength{\tabcolsep}{3.4pt}
\caption{Aggregate Results Over All Delivered Packets.
Latency in ms, Best Value in Each Row in Bold and Second Best Underlined.
Revisits Is the Share of Packets that Visit a Satellite More Than Once.}
\label{tab:aggregate}
\centering
\footnotesize
\begin{tabular}{lrrrrr}
\toprule
 & \multicolumn{2}{c}{RAMA} & \multicolumn{2}{c}{DDQN} & \\
\cmidrule(lr){2-3}\cmidrule(lr){4-5}
 & Frozen & Online & Frozen & Online & Dijkstra \\
\midrule
\multicolumn{6}{l}{\textit{(a) Kepler}}\\
\addlinespace[2pt]
Mean latency              & 46.51 & \underline{45.90} & 53.71 & 186.36 & \textbf{45.20} \\
Median latency           & 43.94 & \underline{43.78} & 45.87 & 67.83 & \textbf{42.76} \\
$90^{\text{th}}$ pct. latency         & \underline{68.24} & \textbf{67.63} & 74.41 & 477.42 & 71.36 \\
$99^{\text{th}}$ pct. latency         & 98.71 & \underline{96.41} & 296.01 & 1{,}636.81 & \textbf{81.86} \\
Latency std.\ dev.        & 24.47 & \underline{22.39} & 50.39 & 320.40 & \textbf{16.65} \\
Latency above $0.5$~s       & \textbf{0} & \underline{10} & 354 & 14{,}780 & \textbf{0} \\
Mean hops         & 7.81 & \underline{7.58} & 8.37 & 27.04 & \textbf{6.80} \\
Revisits          & \underline{0.85\%} & 0.95\% & 2.95\% & 21.59\% & \textbf{0.002\%} \\
\midrule
\multicolumn{6}{l}{\textit{(b) Starlink shell 1}}\\
\addlinespace[2pt]
Mean latency                  & 116.56  & \underline{86.79}   & 1{,}726.22  & 1{,}661.92  & \textbf{57.13} \\
Median latency                & 64.29   & \underline{62.56}   & 312.79      & 842.15      & \textbf{54.85} \\
$90^{\text{th}}$ pct. latency & 251.02  & \underline{146.01}  & 4{,}275.90  & 4{,}238.56  & \textbf{85.71} \\
$99^{\text{th}}$ pct. latency & 791.48  & \underline{484.55}  & 19{,}642.82 & 10{,}651.59 & \textbf{121.16} \\
Latency std.\ dev.            & 159.87  & \underline{92.12}   & 4{,}009.54  & 2{,}225.16  & \textbf{21.85} \\
Latency above $0.5$~s         & 7{,}027 & \underline{1{,}439} & 61{,}386    & 98{,}321    & \textbf{0} \\
Mean hops                     & 40.72   & \underline{30.24}   & 483.08      & 391.12      & \textbf{12.62} \\
Revisits                      & 16.77\% & \underline{12.30\%} & 62.41\%     & 84.71\%     & \textbf{0.11\%} \\
\bottomrule
\end{tabular}\vspace{-20pt}
\end{table}

\Cref{tab:aggregate} gives the aggregate results on both constellations.
On both constellations and in both regimes, RAMA outperforms DDQN on every aggregate, and online RAMA is the best learned method on all but two of them.
On Kepler, the advantage is concentrated among the slowest packets.
Compared to frozen DDQN, frozen RAMA lowers mean latency by 13.4\% and $99^{\text{th}}$-percentile latency by 66.7\% on Kepler.
Meanwhile on Starlink shell 1, frozen RAMA lowers mean latency by 93.2\% and $99^{\text{th}}$-percentile latency by 96.0\% on Starlink.
Frozen RAMA closes 84.6\% of the mean-latency gap between frozen DDQN and Dijkstra on Kepler and 96.4\% on Starlink.
Dijkstra keeps the best value on almost every aggregate but on Kepler RAMA has the lower $90^{\text{th}}$-percentile latency in both regimes.

\begin{figure}[!t]
\centering
\vspace*{-5pt}
\captionsetup[subfloat]{farskip=0pt,captionskip=0pt}
\includegraphics[width=\columnwidth]{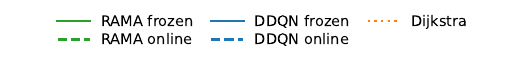}\\
\subfloat[Kepler]{\includegraphics[width=0.88\columnwidth]{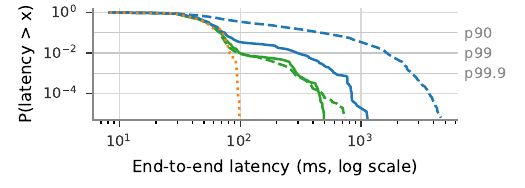}\label{fig:ccdf_kepler}}\\
\subfloat[Starlink]{\includegraphics[width=0.88\columnwidth]{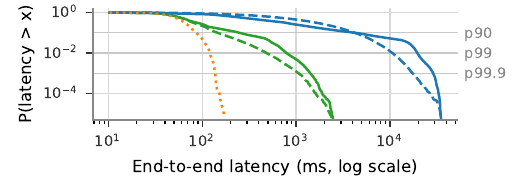}\label{fig:ccdf_starlink}}
\caption{Complementary cumulative distribution of end-to-end latency for RAMA frozen, RAMA online, DDQN frozen, DDQN online and Dijkstra.
}
\label{fig:ccdf}\vspace{-18pt}
\end{figure}

\Cref{fig:ccdf} shows the full latency distributions.
In \Cref{fig:ccdf} (a), RAMA's latency is lower than Dijkstra's at every percentile from about the $84^{\text{th}}$ to the $95.4^{\text{th}}$ when frozen and to the $97.8^{\text{th}}$ when updated online. 
It is higher at every percentile above.
In \Cref{fig:ccdf} (b), RAMA's latency is higher than Dijkstra's at every percentile in both regimes.

\Cref{tab:hopclass} splits each run's packets into long paths and other paths, which decomposes its mean latency into the contribution of each category.
On Kepler, long paths contribute 3.37~ms to frozen RAMA's mean latency and 9.38~ms to frozen DDQN's, which accounts for 6.0~ms of the 7.2~ms difference between their means.
Frozen DDQN's long paths average 39.0 hops, and 76.3\% of them revisit a satellite, against 22.0 hops and 32.6\% for frozen RAMA.
On Starlink, long paths contribute 87.80~ms to frozen RAMA's mean latency and 1,721.05~ms to frozen DDQN's.
This difference of 1,633.25~ms is larger than the 1,609.66~ms difference between their means.
The other paths contribute 23.59~ms more to RAMA's mean latency than to DDQN's, since far more of RAMA's packets take them.
The same pattern can be observed for the online regimes as well.
Thus, avoiding long detours gives RAMA most of its advantage over DDQN on Kepler and all of it on Starlink.

\begin{table}[!t]
\renewcommand{\arraystretch}{1}
\setlength{\tabcolsep}{3.4pt}
\caption{Mean Latency Split Between Long Paths and Other Paths.
Latency in ms.
A Category's Contribution is its Share of Packets multiplied by its Mean Latency, and the Two Contributions Sum to the Mean Latency in \Cref{tab:aggregate}.}
\label{tab:hopclass}
\centering
\footnotesize
\begin{tabular}{lrrrrr}
\toprule
 & \multicolumn{2}{c}{RAMA} & \multicolumn{2}{c}{DDQN} & \\
\cmidrule(lr){2-3}\cmidrule(lr){4-5}
 & Frozen & Online & Frozen & Online & Dijkstra \\
\midrule
\multicolumn{6}{l}{\textit{(a) Kepler, long paths have more than 14 hops}}\\
\addlinespace[2pt]
Long-path share         & 2.55\% & 1.66\% & 3.83\% & 36.26\% & 0\% \\
Long: mean latency      & 132.23 & 142.96 & 245.24 & 425.39  & --  \\
Long: contribution      & 3.37   & 2.38   & 9.38   & 154.24  & 0   \\
Other: mean latency     & 44.27  & 44.26  & 46.09  & 50.39   & 45.20 \\
Other: contribution     & 43.15  & 43.52  & 44.32  & 32.12   & 45.20 \\
\midrule
\multicolumn{6}{l}{\textit{(b) Starlink shell 1, long paths have more than 24 hops}}\\
\addlinespace[2pt]
Long-path share         & 43.40\% & 44.14\% & 89.52\%     & 91.05\%     & 1.31\% \\
Long: mean latency      & 202.32  & 133.01  & 1{,}922.56  & 1{,}819.20  & 114.97 \\
Long: contribution      & 87.80   & 58.71   & 1{,}721.05  & 1{,}656.30  & 1.50   \\
Other: mean latency     & 50.81   & 50.27   & 49.33       & 62.80       & 56.37  \\
Other: contribution     & 28.76   & 28.08   & 5.17        & 5.62        & 55.63  \\
\bottomrule
\end{tabular}\vspace{-20pt}
\end{table}

Updated online, RAMA's performance does not degrade on either constellation while the same does not hold for DDQN.
On Kepler, the mean latency of frozen RAMA, online RAMA, frozen DDQN and Dijkstra, rises by 7.7\% to 11.4\% between packets created in the first and last 300~s of the run, so this drift is not caused by learning.
Online DDQN's mean latency instead rises by 169.3\%, from 63.27~ms to 170.41~ms.
As it updates, it takes more long detours: its share of packets above 14 hops rises from 3.83\% when frozen to 36.26\%, and its revisits from 2.95\% to 21.59\%.
On Starlink, the mean latency of frozen RAMA, online RAMA, frozen DDQN and Dijkstra changes by $-2.2\%$ to $+4.3\%$ between packets created in the first and last 300~s of the run.
Online DDQN's mean latency instead rises by 46.3\%, from 1{,}338.50~ms to 1{,}957.74~ms.
Most of its packets already take long paths when frozen, but as it updates, its revisits rise from 62.41\% to 84.71\%.
Online RAMA instead improves on its frozen performance, lowering mean latency by 25.5\% and $99^{\text{th}}$-percentile latency by 38.8\%.
It does so mainly by shortening its paths above 24 hops, from 73.7 to 48.4 hops on average.

\subsection{Scalability}
\label{sec:scaling}

We first examine how the methods scale from Kepler to Starlink.
From 140 to 1,584 satellites, Dijkstra's mean latency grows by a factor of 1.26 and its mean hop count by 1.86.
RAMA's mean latency grows by a factor of 2.51 and its hop count by 5.22 when frozen, and by 1.89 and 3.99 when updated online.
DDQN's mean latency grows by a factor of 32.14 and its hop count by 57.71 when frozen, and by 8.92 and 14.46 when updated online.
Its online factors are smaller only as online updates had already raised its Kepler mean latency to 3.47 times that of frozen DDQN, while on Starlink both regimes reach similar mean latencies.
RAMA's advantage over DDQN therefore widens on the larger constellation, while its gap to Dijkstra also grows.
Frozen RAMA's $99^{\text{th}}$-percentile latency grows faster than its mean, by a factor of 8.02 against 2.51.
Dijkstra's grows by a factor of only 1.48.
This growth comes with more packets taking long detours and more revisits.
Part of this growth comes from the geometry of Starlink shell 1, where satellites moving north and south pass over the same locations.
On Starlink, 23.4\% of frozen RAMA's packets first come within 2,000~km of $d$ on a satellite moving north while $d$ moves south, or south while $d$ moves north.
From that point, these packets need a median of 20 more hops to reach $d$, against 4 more hops for packets that arrive on a satellite moving in the same direction as $d$.
Although they are less than a quarter of all packets, they account for 32.5\% of frozen RAMA's hops.
On the near-polar Kepler constellation, 15.5\% of packets arrive this way, and they need a median of one more satellite.
Online updates therefore help RAMA most on Starlink, while on Kepler there is only a small decrease in mean latency frozen RAMA already takes few long detours.

\subsection{Trade-Offs}
\label{sec:tradeoffs}
 Dijkstra requires exact knowledge of the current constellation topology at every topology update. In practice, such a scheme must continuously collect network-wide state information and recompute routes over a graph that grows with constellation size, a cost that is not included in our evaluation.
RAMA instead routes from local observations, its own history and its memory bank, at a per-decision cost that is independent of constellation size.
The cost of routing from local information is small on Kepler and larger on Starlink.
On both constellations, RAMA's long detours account for more than the whole gap (\Cref{tab:hopclass}).

Relative to DDQN, RAMA trades additional computation and memory for lower latency. Each decision requires memory retrieval and Transformer-based processing of the current observation, routing history, neighbor information, and retrieved experiences. This incurs higher overhead than DDQN's feed-forward inference on the current observation alone. Under online adaptation, both methods incur the cost of gradient updates, whereas frozen RAMA can continue adapting through memory retrieval and accumulation without further training.

\section{Conclusion}
\label{sec:conclusion}
We presented RAMA, a retrieval-augmented DRL routing framework that equips each satellite with hindsight-labeled episodic memory and Transformer-based history-aware decision making. Evaluations on realistic LEO constellations show that RAMA consistently outperforms DDQN and approaches shortest-path routing using only local information. Its gains arise primarily from eliminating the long, looping detours responsible for DDQN's worst-case packet deliveries, yielding lower tail latency and greater robustness under online adaptation. As constellations grow, RAMA's retrieval-augmented memory offers a scalable means of continually leveraging new experience while maintaining decentralized routing efficiency.

\bibliographystyle{IEEEtran}
\bibliography{references}

\end{document}